\pdfoutput=1
\documentclass[twocolumn,prl,prepreint,preprintnumbers,amsmath,amssymb,floatfix]{revtex4}
\usepackage[pdftex]{graphicx}
\usepackage{color}
\usepackage{dcolumn}
\usepackage{bm}
\usepackage{times}
\usepackage{amsmath}
\usepackage{amssymb}
\newcommand\nn{\nonumber}
\newcommand\bea{\begin{eqnarray}}
\newcommand\eea{\end{eqnarray}}
\newcommand\f{\frac}
\newcommand\p{\partial}
\newcommand\la{\langle}
\newcommand\ra{\rangle}

\begin{document}

\title{Role of conserved quantities in normal heat transport in one-dimenison}
\author{Suman G. Das}
\email{suman@rri.res.in}
\affiliation{Raman Research Institute, CV Raman Avenue, Sadashivanagar, Bangalore 560080, India}
\author{Abhishek Dhar}
\email{abhishek.dhar@icts.res.in}
\affiliation{ International Centre for Theoretical Sciences, TIFR, Bangalore 560012, India}

\vspace{0.5cm}

\date{\today}

\begin{abstract} 
Although one-dimensional systems that exhibit translational symmetry are generally believed to exhibit anomalous heat transport, 
previous work has shown that the model of coupled rotators on a one-dimensional lattice constitute a possible exception. 
We investigate the equilibrium spatiotemporal correlations of energy and momentum of the rotator model, and find that both these fields
diffuse normally. The normal diffusion is explained within the framework of fluctuating hydrodynamics by observing that the angle variables 
do not constitute a conserved field,
which leads to the absence of long-wavelength currents in the system. As an outcome of our analysis, we propose some general criteria
for normal transport based on the existence of conservation laws.

\end{abstract}

\maketitle

The nature of heat transport in one-dimensional systems has been in debate over the last few decades. Most numerical as well 
as analytical studies of momentum-conserving
systems show an anomalous behavior of the heat conductivity $\kappa$ \cite{LLP03,dhar08,LLP97,dhar02,grass02,casati03,mai07,delfini06,narayan,
wang04,pereverez03,lukkarinen,BBO,onuttom03,beijeren,mendl13,spohn13}. 
In the limit of large system size $N$, one generally finds that
\bea
\kappa \sim N^a
\nn
\eea
with $0<a<1$, which means that the thermodynamic limit for the conductivity does not exist and Fourier's law is not valid.
However, there are certain contradicting claims in the literature regarding the generality of this scenario. 
It has been claimed that the value of $a$ may depend on the properties of the inter-particle potential and that one can get normal transport, i.e $a=0$,  in certain kinds of potentials and parameter regimes. 
For example, normal transport has been reported in the so-called rotor model 
at sufficiently high temperatures \cite{giardina00,gendelman00}. 
On the other hand there are other studies on several systems including the $\alpha-\beta$ Fermi-Pasta-Ulam system, which find normal transport at low temperatures  and anomalous transport at high temperatures \cite{zhao12,savin13}. 
Extensive numerical studies of some cases suggest  the possibility
that some of these may be arising from finite-size effects \cite{das14,benenti14}.
The recent approach to transport theory, based on fluctuating hydrodynamics \cite{beijeren,mendl13,spohn13}, offers the possibility of a unifying perspective.
The main aim of this Letter is to provide an understanding of the numerical observations on the rotor model within the framework of 
fluctuating hydrodynamics, which helps illustrate the role of conservation laws in heat transport.

We consider one-dimensional systems on a ring, each particle described by a coordinate $q(x)$ and momentum $p(x)$, with $x=1,\ldots,N$. Defining 
$r(x)=q(x+1)-q(x)$, the Hamiltonian is written as $H = \sum_{x=1}^N e(x)$, where $e(x) = \f{p^2(x)}{2}+ V[r(x)]$. The key ingredient in hydrodynamics
is to identify the locally conserved fields in the system, with their fluctuations around the equilibrium value represented by the vector $\vec{u}(x,t)=\{u_1,\ldots,u_n\}$ for $n$ conserved variables. The corresponding 
currents are denoted by $\vec{j}(x,t)$. In fluctuating hydrodynamics, one adds noise and dissipation terms to the currents, and takes the 
continuum limit to obtain
\bea
\p_t u_\alpha= -\p_x  \left[j_\alpha - \p_x {D}_{\alpha \beta} u_\beta + 
{B}_{\alpha \beta} \xi_\beta \right]~.~~~~\label{EOM}
\eea

The noise terms $\xi$ are stationary Gaussian processes satisfying $\la \xi_\alpha(x,t) \ra =0 $ and 
$\la \xi_\alpha(x,t)\xi_\beta(x^\prime,t^\prime)\ra=\delta_{\alpha\beta}\delta(x-x^\prime)\delta(t-t^\prime)$. 
The noise and dissipation matrices ${B},{D}$, satisfy 
the fluctuation-dissipation relation $ D C^0 + C^0 {D} = {B} {B}^T$, 
where $C^0_{\alpha \beta}= \la u_\alpha(0,0) u_\beta(0,0)\ra$, angular brackets denoting equilibrium averages. 
The introduction of the noise and dissipation terms rules out integrability
indirectly, and also allows one to derive and solve the corresponding Fokker-Planck equations with a plausible ansatz \cite{spohn13}. 
But quite apart from the mathematical convenience, it is important to understand what they physically represent. In deriving the continuum
hydrodynamic equations, one necessarily loses information about the short-wavelength behavior, with the continuum equations containing
only the modes that relax on the longest length and time scales. Although a rigorous derivation is lacking, the fluctuation and 
dissipation terms may be taken to represent the effects of the faster modes on the long-wavelength behavior of the system. The fluctuation-dissipation
relation asserts that these modes are thermalized, whereas the lack of correlation of the noise terms in space and time indicate their fast 
relaxation in equilibrium. It should be possible in principle to derive the diffusion coefficients in Eq.~(\ref{EOM}) following 
the Hamiltonian
derivation of the Navier-Stokes equations in \cite{sasa14}.

In a general one-dimensional non-integrable system, one has three conserved fields and corresponding currents given respectively by
$\vec{u}=(r,p,e)$ and $\vec{j}=(-p,P,pP)$, where the field now represents the local fluctuations of stretch, momentum and energy about their 
respective equilibrium values. The local pressure field is 
$P(x,t)\equiv -\partial_r V(r(x,t))$. The thermodynamic pressure is 
given by $\bar{P}=-\la \partial_r V(r) \ra$.
This is the case addressed in \cite{spohn13}, where the hydrodynamics is treated by expanding the currents up to second order in the fields:
\bea
j_\alpha=    A_{\alpha \beta} u_\beta + H^\alpha_{\beta \gamma}  u_\beta u_\gamma~.~~~~\label{current}
\eea 
The tensors $A$ and $H$ are expressed in terms of the derivatives of the pressure $\bar{P}$ with respect to the energy and volume of the corresponding
microcanonical ensemble, using Maxwell-type thermodynamic relations. When the expansion 
is truncated to first order (i.e set $H^\alpha_{\beta \gamma}=0$), the resulting linear hydrodynamics 
can be diagonalized to obtain the three normal modes $\phi_\alpha$, one of which is a stationary heat mode and 
the other two are sound-modes traveling at speeds $\pm c$. 
The second-order terms in the expansion
can then be expressed as a coupling between these normal modes $\phi_\alpha$,  and the dynamical correlators 
$ \la \phi_\alpha(x,t) \phi_\beta(0,0) \ra$ of the resulting theory are calculated within a mode coupling approximation. 
In particular one finds that the heat mode correlations exhibit superdiffusive (Levy) scaling, which
implies super-diffusive transport of heat and universal non-zero values for $a$. The predicted exponents are well-verified
numerically for hard-particle type gases and classical anharmonic chains \cite{mendl14,das14b}. Scaling predictions from \cite{beijeren}
are also found to hold for solutions of the Gross-Pitaevskii equation applied to the long-wavelength behavior of 
one-dimensional Bose gases \cite{kulkarni13}.

Thus to explore the possibility of normal diffusion of heat, one has to look for cases in which there are less than three conserved fields.
Since energy is necessarily a constant of motion, the only possibilities are where either the stretch $r(x)$ or the momentum $p(x)$ (or both) 
are not conserved. We claim that any of these cases lead to normal transport of energy. The case of non-conservation of momentum has been treated 
quite extensively through several models in the literature, and thus we focus primarily on the case where stretch is not conserved. However, as 
we shall see later, the hydrodynamics of the energy mode is practically identical for the two cases.

To explore the case of non-conserved stretch, consider a Hamiltonian $H$ with $V(r)=V_0\cos(r)$, 
commonly known as the coupled rotator (CR) model. Convincing numerical evidence 
\cite{giardina00,gendelman00} 
exists for a finite thermal conductivity in this model, based on direct molecular dynamics 
simulations of the steady state as well as equilibrium current autocorrelations.
Note that since the potential is periodic, the angle variables $r(x)$ should be taken to lie within an appropriate finite range to ensure that
the partition function is finite and therefore the canonical ensemble is well-defined. For our analysis we restrict it within the
interval $(0,2\pi)$, although any range that is an integer multiple of $2\pi$ should be expected to have identical macroscopic properties.
The equation of motion of $r(x)$ within the range $(0,2\pi)$ is  $\dot{r}(x)=p(x+1)-p(x)$,
but since $r$ is restricted within the cell, one should add boundary terms at 
$0$
and $2\pi$ such that the $r(x)$ are reset whenever the prescribed range is exceeded. 
This discontinuity in $r(x)$ is not in general compensated by a simultaneous change in the nearest neighbors,
which means that $r(x)$ is not a locally (or globally) 
conserved field. 

\begin{figure}
\includegraphics[width=3.5in,height=2in]{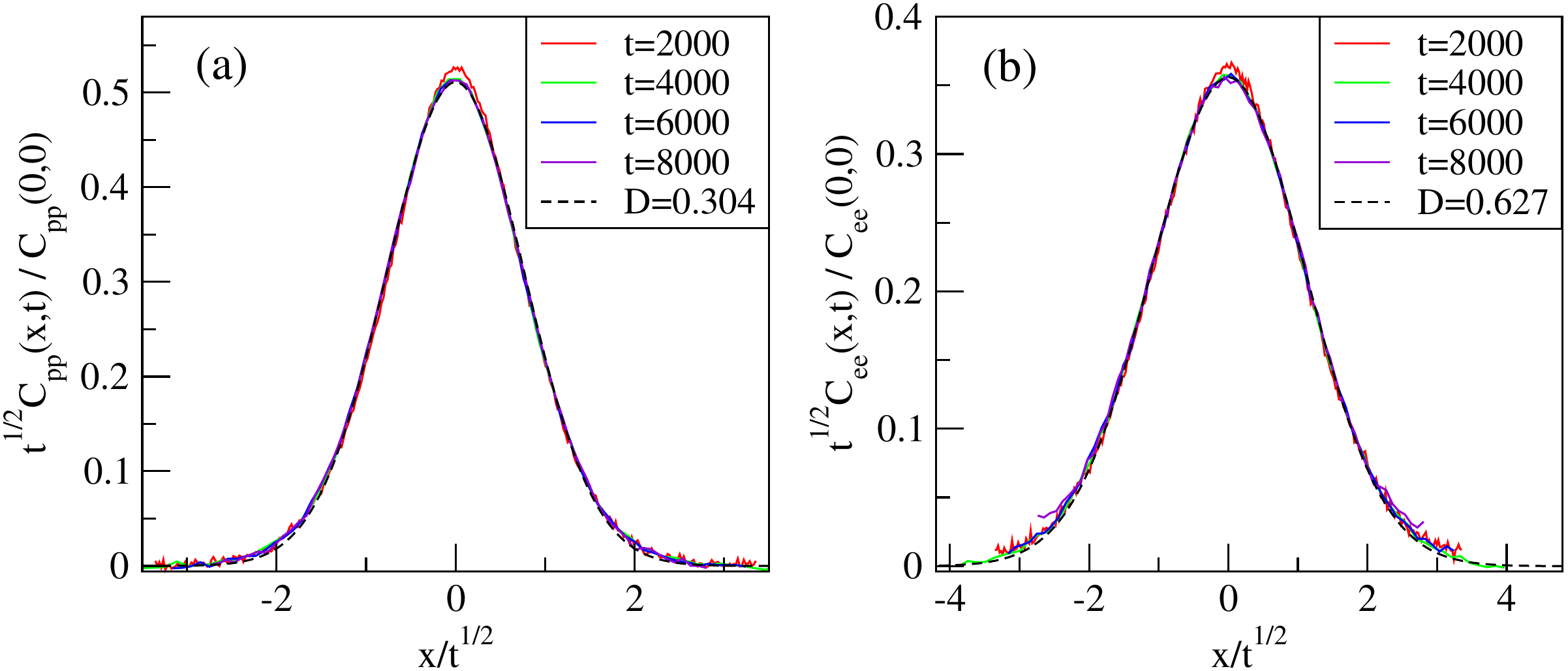}
\caption{(color online) Plot of the autocorrelation of momentum (left) and energy (right), scaled in accordance with normal diffusion and normalized 
with $C(0,0)$,
which is computed numerically from the equilibrium distribution. The dashed black  lines correspond to 
normalized Gaussians with the respective diffusion constants mentioned in the figures.} 
\label{scaling}
\end{figure} 

\begin{figure}
\includegraphics[width=2.5in]{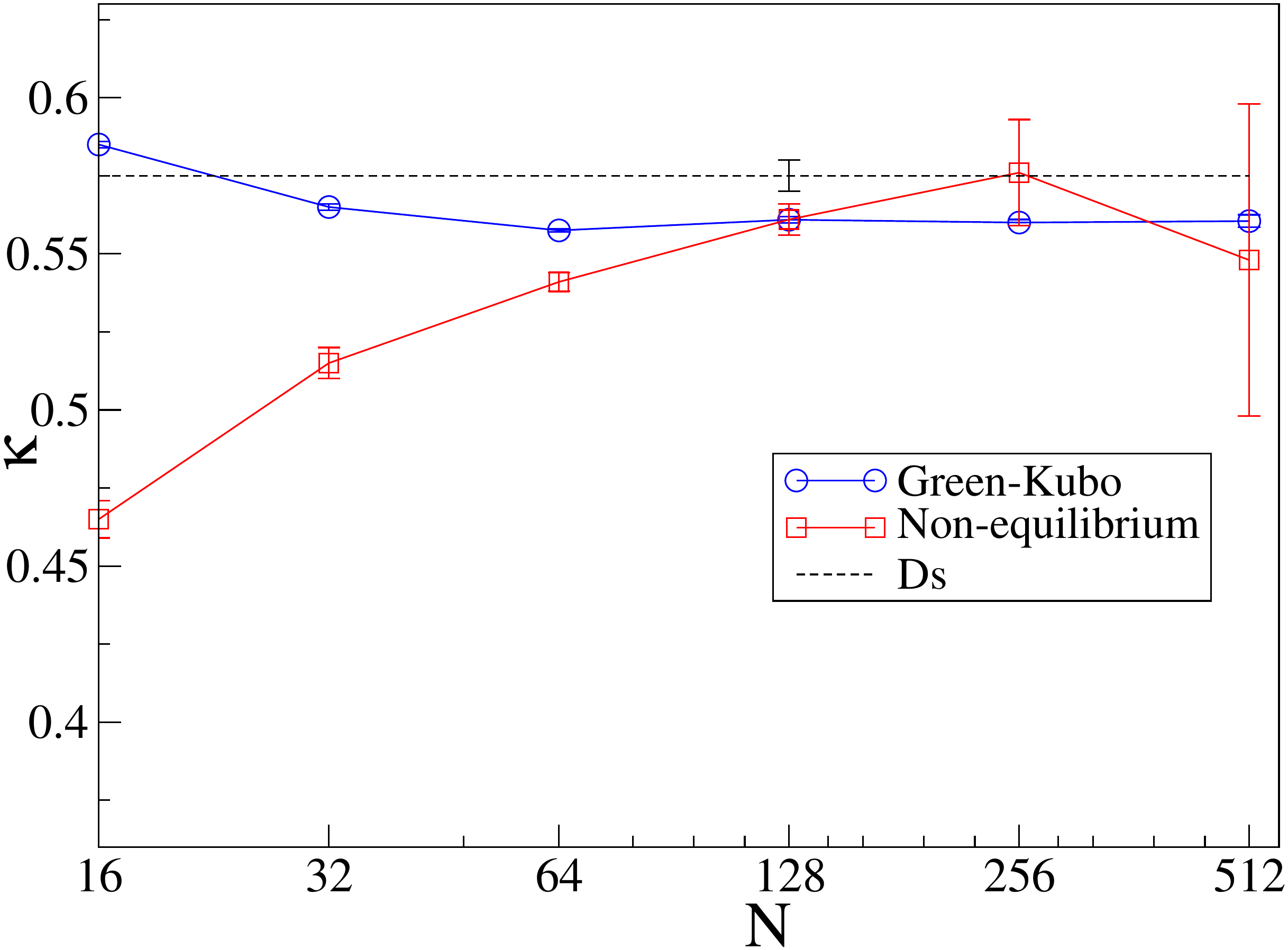}
\caption{(color online) Plot of conductivity for $V_0=1$ and $T=1$ through three different methods. The circles and the boxes show the size-dependent behavior of the 
conductivity as obtained from the Green-Kubo formula and direct nonequilibrium simulations respectively. The dashed straight line
is the value of the conductivity as predicted from the specific heat density and the diffusivity of the heat mode. { The small discrepancy between
the diffusivity and the Green-Kubo results is likely due to the fact that terms second-order in the temperature gradient are ignored in the derivation of
the expression for diffusivity.}}
\label{conductivity}
\end{figure} 

This crucial observation implies that the thermodynamic conjugate of stretch, the pressure ${\bar{P}}$, is
necessarily zero.  In the case of stretch-conserving models, a non-zero pressure is incorporated into the equilibrium description by modifying the 
Gibbs 
weight of each microstate such that  
$Prob(\{r(x)\}) \sim \Pi_x~e^{-\beta[V(r(x))+\bar{P} r(x)]}$. This measure remains invariant under the dynamics since $\sum_x r(x)$ is conserved. 
However,
when the stretch is not conserved, as in the case of the CR model, the measure is invariant only for $\bar{P}=0$. It is important to emphasize that 
the equilibrium pressure of the rotator model is {\it identically} zero, unlike the special case
of zero pressure in stretch conserving systems, where the derivatives of pressure with respect to energy and volume are finite even at zero 
pressure and thus the normal modes remain coupled \cite{spohn13}. For the CR model, the  non-conserved field $r$ 
drops out from the hydrodynamic description.
Since $\bar{P}$ is identically zero, the coupling tensors $A$ and $H$  in Eq.~(\ref{current}) 
(and any higher-order terms that may be included) vanish, and
the hydrodynamic currents $j$ are zero. The local pressure continues
 to fluctuate on a microscopic scale, but these fluctuations are effectively incorporated into the noise and dissipation terms in the net current.
Thus the conserved field in this model is $\vec{u}=(p,e)$, and the corresponding hydrodynamic equations are
\bea
\p_t u_\alpha= -\p_x  \left[- \p_x {D}_{\alpha \beta} u_\beta + 
{B}_{\alpha \beta} \xi_\beta \right]~.
\label{diffusion}
\eea
As in the case of stretch-conserving systems, we find numerically that the cross-correlations decay rapidly, and then the 
auto-correlations $C_{\alpha \alpha}(x,t)\equiv \la u_\alpha(0,0) u_\alpha(x,t)\ra$ can be shown to be
\bea
C_{\alpha \alpha}(x,t)=\f{1}{\sqrt{4 \pi D_{\alpha \alpha} t}} \exp~\left[-\frac{x^2}{4 D_{\alpha \alpha} t}\right]~,
\label{corr}
\eea
implying diffusive transport of both momentum and energy. Sound modes are absent in the hydrodynamic limit.

We check these predictions with numerical simulations. We have performed molecular dynamics simulations of the CR model in equilibrium, 
choosing the initial conditions from the Gibbs distribution, and integrating the system using
the fourth order Runge-Kutta as well as the velocity Verlet algorithm. 
The normalized correlation functions of momentum, $C_{pp}(x,t)/C_{pp}(0,0)$ and energy, $C_{ee}(x,t)/C_{ee}(0,0)$ are shown in Fig.~(\ref{scaling}), 
for $V_0=1$ temperature $T=1$ and a system of size $N=400$.  
The momentum and energy modes have been scaled diffusively and show excellent collapse. In fact, the scaled correlation functions are 
fitted very well 
by Gaussian functions (dashed curves), showing that the correlation functions are as predicted in Eq.~(\ref{corr}). Note that the form 
of these correlation functions are completely different from the ones in systems with three conserved fields, where one finds 
a non-Gaussian central peak in addition to traveling sound peaks. 
We must mention that
similar diffusive nature of momentum correlatios in the rotator model have recently been reported in \cite{li2014}.

Using Fourier's law and { assuming that temperature fluctuations are small}, 
it can be shown that $\kappa=D s$, where $D$ is the energy diffusivity and $s$ is the specific heat density. 
For the present system with $V_0=1$ it is 
calculated that the partition function 
$Z=\int_{-\infty}^\infty dp~ e^{-\f{1}{2}\beta p^2}\int_0^{2 \pi}  dr~ e^{-\beta \cos (r)} =\sqrt{\f{2\pi}{\beta}} I_0 (\beta)$,
where $I_n(x)$ is 
the $n$-th modified Bessel function of the first kind. From this one finds that 
\bea
s\equiv \f{1}{T^2}\f{\p^2}{\p \beta^2} ln Z =\f{1}{2}\left[1+\beta^2(1+\f{I_2(\beta)I_0(\beta) - 2{I_1(\beta)}^2}
{I_0(\beta)^2})\right].
\nn
\eea
For our parameters, $s=0.9168$, and using $D$ as determined from the numerical fitting in Fig.~(1b), we get
$\kappa=0.5749$.

\begin{figure}
\includegraphics[width=3in,height=2.2in]{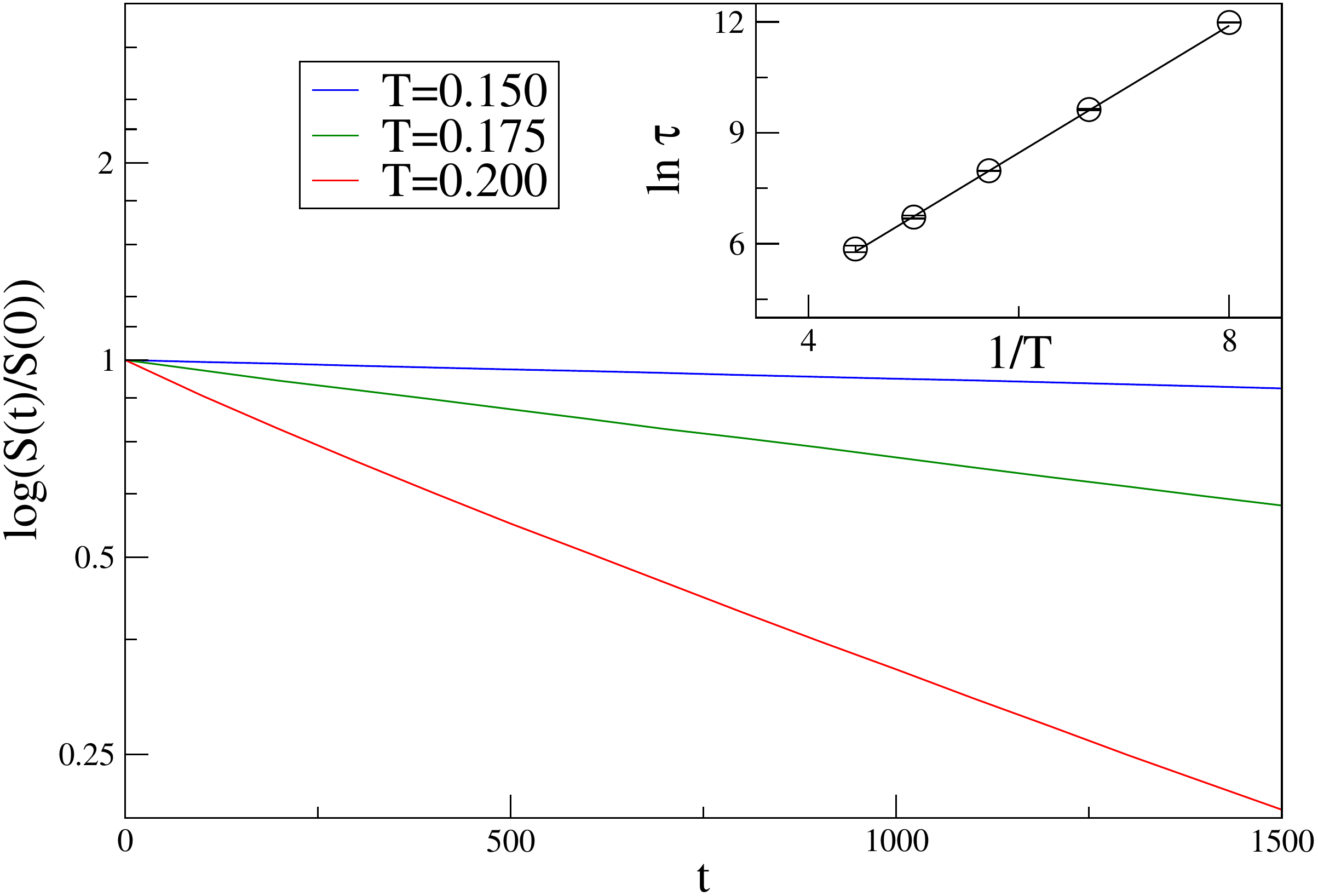}
\caption{(color online) Plot of $S(t)$ at various temperatures for the CR model with $V_0=1$.
Note that the decay is exponential, and even at the lowest temperature a slow
decay is evident. The inset shows the logarithm of the correlation time against $1/T$, which appears linear. 
The straight line fit is $y=1.72x-1.86$, and the slope is not too different from $\Delta E = 2$ at this level of approximation
.} 
\label{thetasum}
\end{figure} 

We compare this value with the conductivity obtained directly through two other independent ways. 
When the conductivity is finite, then  
from the Green-Kubo formula it is given by
\bea
\kappa=\lim_{\tau \to \infty}\f{\langle Q_\tau^2 \rangle}{2N T^2\tau},
\nn
\eea
where $Q_\tau=\int_0^\tau J(t) dt$, and the total heat current $J=\sum\limits_{x=1}^N p(x) {\partial V}/{\partial r(x)}$. 
The average is over the equilibrium ensemble at temperature $T$.
A second way of determining $\kappa$ is through direct non-equilibrium simulation of the heat current. Two ends of the rotor chain are connected to 
Langevin baths
at temperatures $T \pm\Delta T$, and the resulting steady-state heat current is calculated and the conductivity is obtained from the standard 
definition. In both these cases, $\kappa$ is computed from simulations performed for different of system sizes. In Fig.~\ref{conductivity}, 
we see that the conductivity from the Green-Kubo formula converges to the value $\kappa\approx 0.56$,  
in close agreement with the value predicted from the diffusivity above. The values from the non-equilibrium simulations are similar, but
the larger error bars make a precise comparison difficult.
\linebreak

To understand the non-conservation of stretch more clearly, we notice that for any conserved field $u$ with zero mean, 
the following exact sum rule holds:
\bea
\sum_x \la u(0,0)u(x,t) \ra = \sum_x \la u(0,0)u(x,0) \ra=\la u^2 (0,0) \ra,
\nn
\eea
where the last equality is for initial conditions with a product measure. 
Since $r(x)$ is not conserved in the CR model, we expect $S(t)\equiv \sum_x \la r(0,0)r(x,t) \ra$ to decay with time 
and go to zero at long times. Since the non-conservation
is a consequence of periodicity, the time-scale associated with the decay of $S(t)$ must be the time-scale required for the variables 
$r_i$ to reach the boundary of the interval $[0,2\pi]$. 
{ In an approximate single-particle picture, it is the time required for a Brownian 
particle to escape the 
potential well represented by one wavelength of the potential}. In the limit of low temperature, this time-scale approaches the Kramers escape time
$\tau \sim \exp(\Delta E/T)$, where $\Delta E$ is the height of the potential well. In other words, at 
low temperatures we expect $S(t) \sim \exp(-t/\tau)$,
with $\tau$ roughly proportional 
to $\exp(2/T)$. Fig.~\ref{thetasum} shows the exponential behavior of $S(t)$, and the inset shows that the decay time indeed scales with $T$ approximately as 
predicted
(see caption for details). 
It was reported in \cite{gendelman00} that there exists a transition from anomalous to normal diffusion in the CR model between $T=0.2$ and $T=0.3$. 
However, as seen in
Fig. \ref{thetasum}, $S(t)$ continues to decay at temperature $0.2$ and below, but the very slow rate of decay means that it is 
hard to numerically 
observe the asymptotically converging thermal conductivity on numerically accessible scales. 
So the apparent anomalous diffusion reported in \cite{gendelman00} 
is possibly a finite-size effect. Similar objections to the claim was made in \cite{giardina00,yang}, 
though not from the viewpoint of conservation laws.

\begin{figure}
\vspace{1cm}
\includegraphics[width=3.4in,height=2.5in]{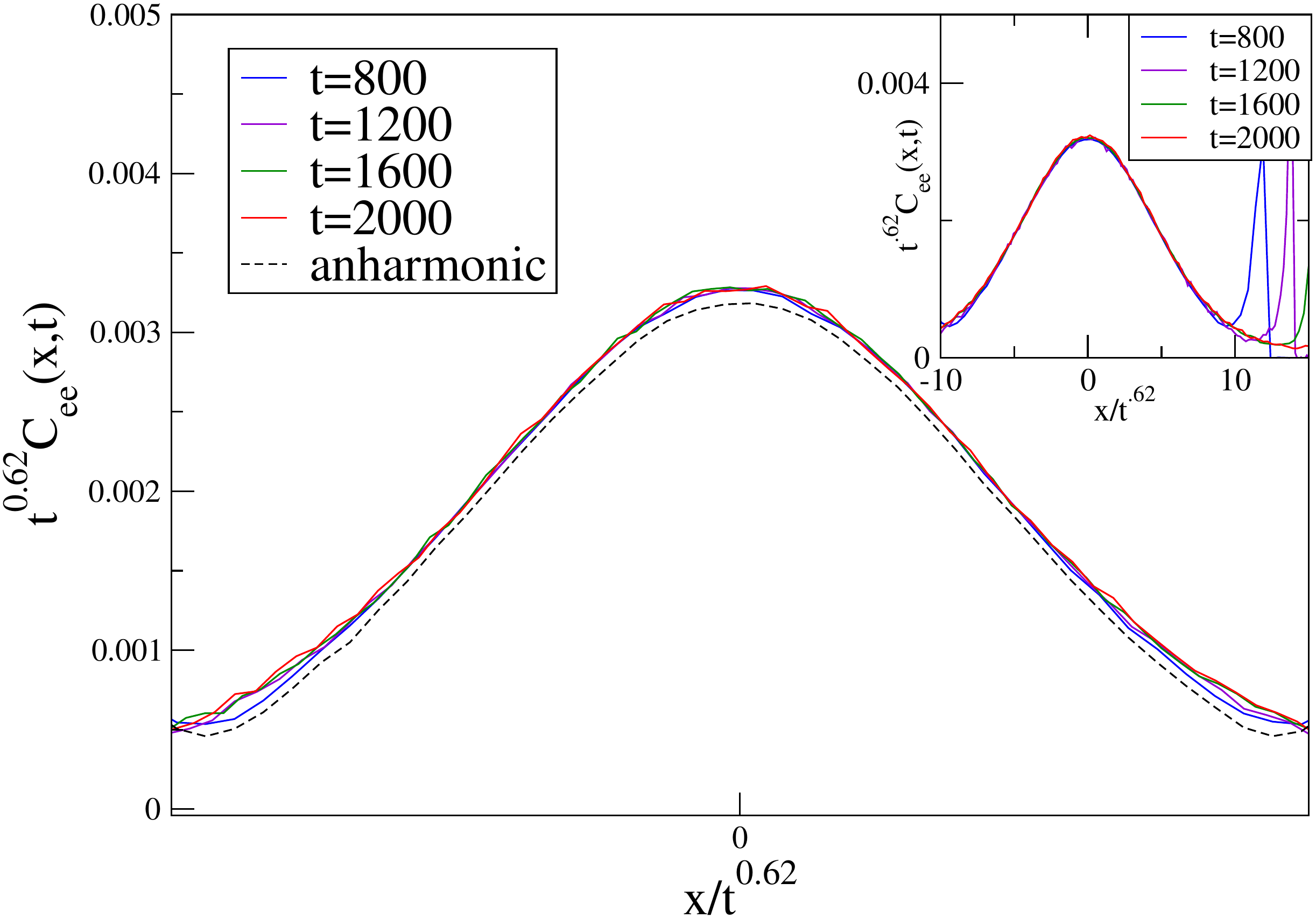}
\caption{(color online) Scaled plots of $C_{ee}(x)$ for the CR model, for $T=0.2$.  
The dashed line is for the corresponding
anharmonic model, at $t=800$. Including higher order terms in the anharmonic potential should produce an even better approximation. 
The inset shows the corresponding plots for the anharmonic model with the same scaling.
} 
\label{heatscl}
\end{figure} 

At low temperatures, each angle variable spends most of its time fluctuating near the potential minimum. Thus at
short times, the hydrodynamics must be well-approximated by an anharmonic expansion of the CR potential
around the minimum at $r=\pi$, while the crossover to normal diffusive behavior should { start to occur} at the time-scales indicated
in Fig.~(\ref{thetasum}). To check this, we consider the anharmonic chain obtained after truncating the CR potential (with $V_0=1$)
at the first stable anharmonic term, obtaining $V(r)=r^2/2-r^4/24+r^6/720$. This system has three conserved fields, as discussed before.
For anharmonic chains at zero pressure, hydrodynamics predicts that the heat mode scales as $x\sim t^{2/3}$ \cite{spohn13}. We simulate the
equilibrium correlations for this model for comparison.
The energy mode for the CR model (Fig.~\ref{heatscl}) satisfies an anomalous scaling 
$x\sim t^{0.62}$. 
The dashed line gives numerical results from the truncated anharmonic potential, which is in good agreement with
the CR model up to the simulation time. The inset shows the same correlations for the anharmonic potential with the same scaling.
The scaling exponent $0.62$ is close to but not yet identical with the prediction in \cite{spohn13}, and is presumably due to the slow separation
of heat and sound modes at low temperatures as seen in simulations elsewhere \cite{das14b}.

\begin{figure}
\vspace{1cm}
\includegraphics[width=3.5in,height=2.5in]{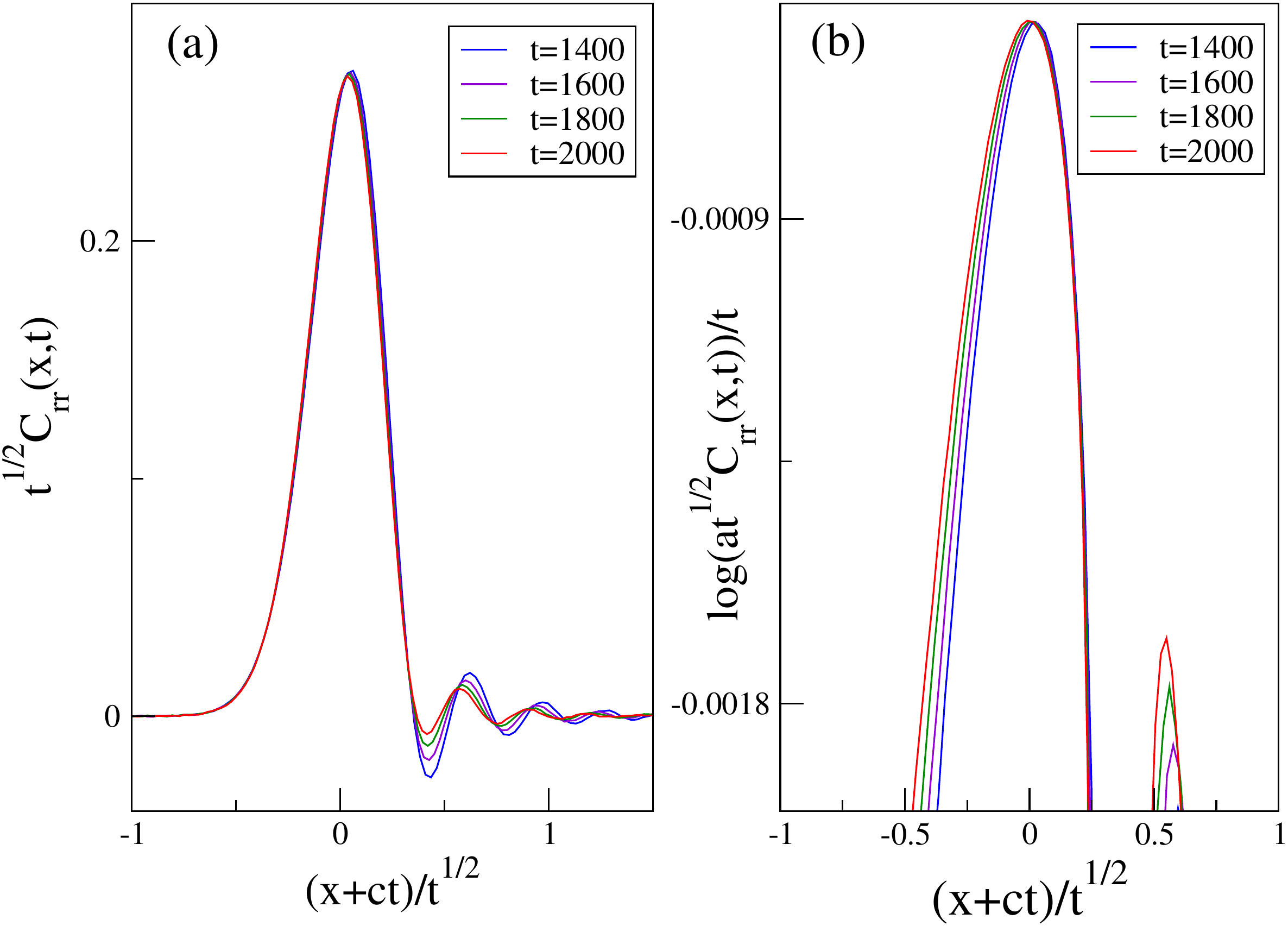}
\caption{(color online) Plot of $C_{rr}(x,t)$ around the left-moving sound peak for $T=0.2$. (a) The anharmonic model correlations scaled with $x\sim t^{1/2}$. 
(b) The CR model scaled similarly with an additional exponentially decaying pre-factor. The sound speed at this temperature is $c=0.938791$. The constant
$a=3.49$.}
\label{soundscl2}
\end{figure}

The stretch-stretch correlations for the anharmonic and CR models agree near the sound peak at short times (not shown). At longer times, $C_{rr}(x,t)$
for the anharmonic chain (Fig. \ref{soundscl2}a) continues to satisfy a $x \sim t^{1/2}$ scaling for the sound mode
as predicted
from nonlinear fluctuating hydrodynamics, but
the sound peaks of the CR model (Fig.~\ref{soundscl2}b) { start to diverge from the anharmonic model}. The scaling in the figure indicates that
close to the peak the sound mode scales as $C_{rr}(x,t)\sim exp(-t/\tau)\f{1}{t^{1/2}}f(x/t^{1/2})$, with $\tau\approx 1880$, 
which is considerably different from $\tau\approx 700$ for $S(t)$ as shown in Fig.~\ref{thetasum}. This is not surprising since this
scaling form does not hold away from peak, as is evident from Fig.~\ref{soundscl2}b.
The exponential term in the sound peak indicates that even at low temperatures 
the sound modes are transient, { but the large $\tau$ means that one would have to wait very long before the sound modes disappear
and a diffusive scaling of the heat mode is observed.}

For momentum non-conserving models, one expects the roles of the stretch and momentum fields to become interchanged. The momentum correlations
would be short-ranged, and consequently the hydrodynamic currents for stretch and energy fields would vanish, leading to normal diffusion
of these fields. Indeed for the harmonic model with random velocity flips (which conserves energy and density but not momentum), a rigorous 
derivation of the hydrodynamic limit (see Sec.~(3) of \cite{bernardinlebowitz}) gives macroscopic diffusion equations similar to 
Eq.~(\ref{diffusion}),
with the field index running over stretch and energy fields. 

It has so far been believed that breaking translational symmetry is crucial to normal heat conduction, but the CR model had remained a puzzle.
We show that it is not translational invariance but the number of conserved fields that decides whether transport is normal.
To put it concisely, {\it whenever stretch (momentum) is not conserved
in a one-dimensional model, the momentum (stretch) and energy fields exhibit normal diffusion}. The possibility of normal transport 
in one dimension has been a matter of long-standing debate, and in this work we identify sufficient hydrodynamic criteria
for normal transport of energy in one-dimensional systems.

Note: During preparation of this manuscript we came to know of a related work on the rotator model \cite{spohn2014} with similar conclusions as ours.

\end{document}